\documentclass[
superscriptaddress,
showpacs,
 amsmath,amssymb,
 aps,
pre,
twocolumn]{revtex4}

\usepackage{graphicx}
\usepackage{dcolumn}
\usepackage{bm}

\begin{document}


\title{Order, intermittency and pressure fluctuations in a
system of proliferating rods}

\author{Sirio Orozco-Fuentes}
\email{sirioanel@fisica.unam.mx}
\affiliation{Instituto de F\'\i sica, Universidad Nacional Aut\'onoma de 
M\'exico, D.F. 04510, M\'exico}
\author{Denis Boyer}
\email{boyer@fisica.unam.mx}
\affiliation{Instituto de F\'\i sica, Universidad Nacional Aut\'onoma de 
M\'exico, D.F. 04510, M\'exico}
\affiliation{Centro de Ciencias de la Complejidad, Universidad Nacional 
Aut\'onoma de M\'exico, D.F. 04510, M\'exico}

\date{\today}

\begin{abstract}
Non-motile elongated bacteria confined in two-dimensional open 
micro-channels can 
exhibit collective motion and form dense monolayers with nematic order if the cells 
proliferate, {\it i.e.}, grow and divide. Using soft molecular dynamics 
simulations of a system of rods interacting through short range mechanical 
forces, we study the effects of the cell growth rate, the cell aspect ratio 
and 
of the sliding friction on nematic ordering and on pressure fluctuations 
in confined environments. Our results indicate that rods with aspect 
ratio $>3.0$ reach quasi-perfect nematic states at low sliding friction. 
At higher frictions, the global nematic order parameter shows intermittent 
fluctuations due to sudden losses of order and 
the time intervals between these bursts are power-law distributed. 
The pressure transverse to the channel axis can vary abruptly in 
time and shows hysteresis due to lateral crowding effects. The 
longitudinal pressure field is on average correlated to nematic order, but 
it is locally very heterogeneous and its distribution follows an inverse 
power-law, in sharp contrast with non-active granular systems. We discuss 
some implications of these findings for tissue growth.
\end{abstract}

 \maketitle

\maketitle


\section{\label{sec:intro}Introduction}

Active suspensions of bacteria or other motile 
particles commonly exhibit collective 
motion and rich nonequilibrium structures at the hydrodynamic scale, such 
as swarming \cite{libchaber,couzin,volfsonprl}, 
instabilities \cite{joanny}, turbulent vortical flows \cite{turb1,turb2,turb3}, 
jamming \cite{jamming} or
aggregation in clusters with giant number fluctuations 
\cite{swinney,peruaniprl}. 
Systems of self-propelled rods that interact through short range mechanical 
forces may provide minimal models for describing colonies of active 
elongated particles \cite{baskaranprl2008,baskaranpre2008,chate,aranson}. 
Despite of the fact that such models ignore chemotaxis and other 
biological signaling processes that may occur in real cell colonies, they are 
thought to be relevant at high cell densities and have actually been able 
to account quantitatively for many experimental observations. For instance, 
rich dynamical features can emerge in active 
rod models with only varying the density and the particle aspect 
ratio \cite{turb1,turb2,gianttheor}.

Whereas most research on active matter has considered motile particles,
the effects of cell proliferation on collective motion are less understood. 
Here we investigate the dynamics of colonies of non-motile but growing and 
dividing rods. Such systems are relevant to
the formation or renewal of biofilms and tissues, and their study may help
to understand the role played by physical constraints during collective cell 
processes such as the growth of a column of hydra \cite{hydra},
tissue growth and repair \cite{tissuebiophys,tissuerepair} or tumor 
growth \cite{tumorjtb}.
Even in the absence of self-propulsion of individual cells, 
cell proliferation generates motion due to excluded 
volume effects, which, in combination with cell anisotropy, can lead to 
nematic ordering and coherent flow patterns \cite{volfsonpnas,physbiol,groisman}.
An important difference with the self-propelled case is that density is 
no longer a control parameter since the system typically self-organizes 
into dense states, starting from a small number of initial cells. 
In addition, as
expansive flows are often generated during growth, pressure gradients can 
be high and may trigger secondary instabilities particular to these
systems \cite{physbiol}. 

\begin{figure*}
\begin{center}
\includegraphics[width=0.87\textwidth]{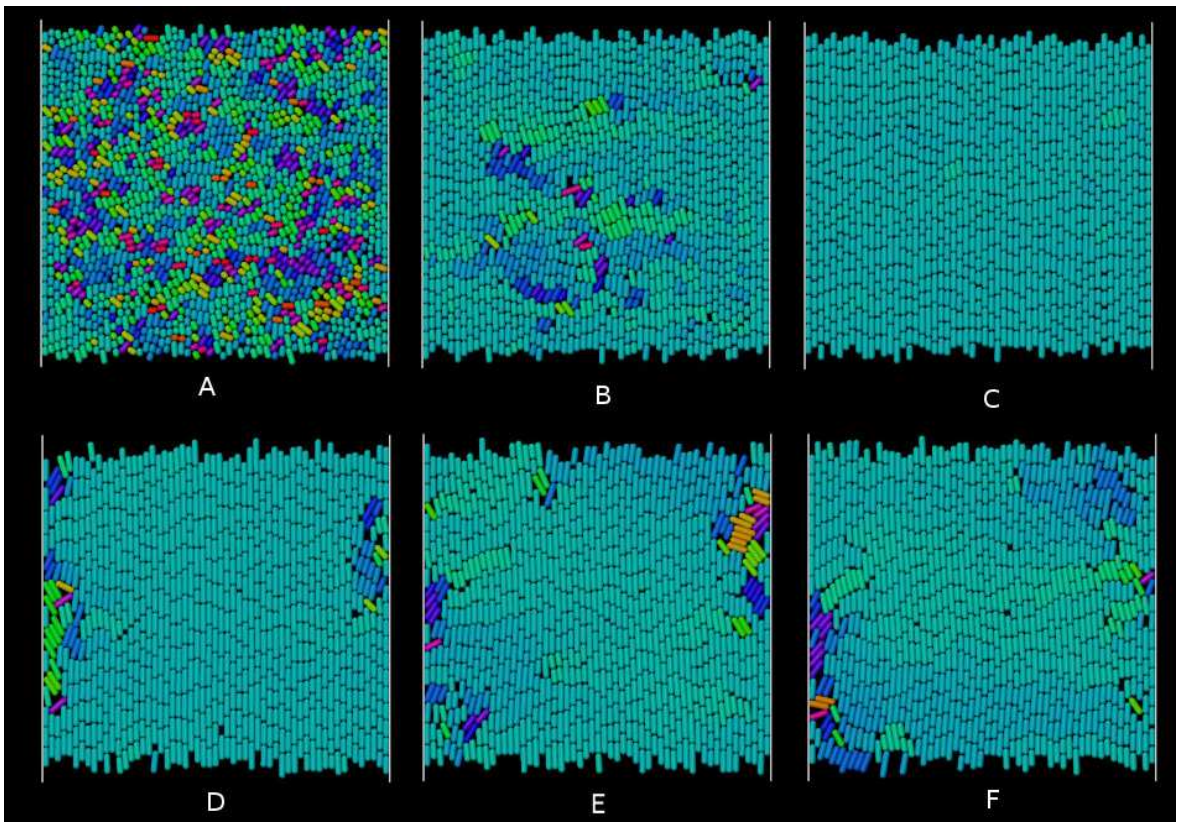}
\caption{(Color online) Typical configurations of simulated bacterial colonies 
filling the channel, with the rods colored according to their orientation. 
{\bf A-C}: Systems of low drag friction ($\mu=10^{-6}$) and different rod 
aspect ratio: $\ell_0=2.0$ (A), $\ell_0=3.0$ (B) and $\ell_0=4.0$ (C). 
The growth rate is $a=1.0$. 
Systems of longer rods have higher nematic order. {\bf D-F}: 
Systems with $\ell_0=4.0$ and of varying drag friction: $\mu=0.45$ (D), 
$\mu=0.50$ (E) and $\mu=0.55$ (F), where disordered regions appear 
intermittently.}
\label{mu045}
\end{center}
\end{figure*}

In this paper, inspired by experiments performed with a non-motile 
strain of {\it E. coli} bacteria (division time $\sim20$ min) in microfluidic 
devices \cite{volfsonpnas}, we perform molecular dynamics simulations of a 
system of growing and dividing rods with repulsive interactions.
This system is confined by the lateral walls of a two-dimensional channel 
of finite length and open at both ends, where the particles can exit 
the channel.
In the microfluidic experiments, the channel was limited in 
the third dimension by two walls whose separation distance was barely larger 
than the diameter of one bacteria. Therefore, although the rods are 
three-dimensional objects in the model, they form a single layer and
their motion is assumed to be two-dimensional. 

As shown by continuum theories of self-propelled particles
\cite{baskaranprl2008,bartolo}, the effect of boundaries and confinement 
have a strong impact on the ordering of active flows, where, for instance, 
the presence of the walls can induce a non-zero polarization. Similarly here, 
non-motile elongated particles push each other while they grow and tend to 
align parallel to the walls of the channel. In the long
time regime, the rods that flow out of the channel are constantly replaced 
by new rods which form a dense model tissue inside the channel, 
with relatively small local density 
fluctuations \cite{volfsonpnas}. In ref. \cite{physbiol}, it was shown with 
the use of a phenomenological continuum theory and discrete element 
simulations that the perfectly ordered active nematic state 
was unstable with respect to small perturbations when a friction 
parameter exceeded a threshold value. This instability is analogous to a 
buckling instability and provokes the growth of the angles between the 
rods and the channel axis, allowing the release of the high 
compressive stresses generated by fully ordered configurations.

The aim of the present study is to investigate numerically the partially 
disordered states formed by these confined proliferating systems in the
long time regime, when the statistical properties of the flow do not depend 
on time. We first quantify the effects on nematic ordering of the rod aspect 
ratio and of the friction that opposes rod motion. We then show that the 
nematic order parameter exhibits intermittent dynamics at intermediate
frictions. We next focus on how the diagonal stress components fluctuate 
in time and space. We find that configurations subjected to larger 
longitudinal stresses are more ordered on average, whereas, locally, the 
distribution of contact forces is very heterogeneous and follows a 
power-law distribution in most cases.

\section{Model description}\label{model_description}

In our approach, thermal noise is neglected and rod dynamics is 
essentially deterministic.
The discrete element soft-particle model used in this paper was described in 
previous works (see, {\it e.g.}, \cite{volfsonpnas} and 
\cite{physbiol}). Briefly, each cell is represented as a rigid rod 
consisting of a cylinder of fixed diameter set to unity for convenience 
and of two hemispherical 
caps at its ends. The length $l(t)$ of a given rod grows exponentially 
at a certain rate $a_g$ and the rod divides in two collinear rods 
of equal lengths
when $l(t)$ reaches an assigned maximal length, denoted as $\ell_m$. 
To avoid spurious synchronization of cell divisions across the population,
the division length $\ell_m$ is chosen randomly at the birth of
each cell from a narrow normal distribution centered at a certain 
value $2\ell_0$ and with standard deviation $0.2\times 2\ell_0$. 
Similarly, $a_g$ is chosen from a distribution centered at a value $a$ 
and with standard deviation $0.1a$. Therefore, $a$ represents 
the mean growth rate of the rods in the system and $l_0$ the mean rod
length at birth. 

The rods are confined in a channel composed of two parallel walls separated 
by a distance $L_x$ (the transversal unit vector is denoted as 
$\hat{x}$) and length $L_z\equiv 
2L$ (the longitudinal unit vector is denoted as $\hat{z}$). We set 
$L_x=L_z=55$ in the following. The rods
cannot form more than one layer in the $\hat{y}$ direction, therefore
motion is bidimensional. The
channel boundaries at $z=\pm L$ are open: when the center of mass of a rod 
crosses one of the boundaries, the rod is removed from the system.

The normal contact
forces between rods are obtained with the Hertzian model applied to
the overlap of virtual spheres centered at the nearest points on the
axes of interacting spherocylinders; similarly, the tangential 
(frictional) forces are given by the dynamic Coulomb friction
($\nu_{cc}$ is the coefficient of friction between cells) 
\cite{volfsonkudrolli}. The microscopic parameters 
characterizing the elastic and dissipative properties of the 
cells coincide (unless indicated) with the ones used in 
\cite{volfsonpnas,physbiol}.
In the experiments of ref.\cite{volfsonpnas}, each cell is also subjected 
to forces due to the surrounding fluid and to the horizontal
walls of the microfluidic chamber. These forces are modeled here by a Stokian
drag force:
\begin{equation}
{\mathbf F}^{(i)}_f=-\mu m^{(i)} {\mathbf v}^{(i)},
\end{equation} 
with ${\mathbf v}^{(i)}$ and $m^{(i)}$ the velocity and mass of rod $i$,
respectively, and $\mu$ a drag friction constant. The contact and friction 
forces above are then used to compute the motion of each rod by integrating 
Newton's equations.

For systems of rods that are perfectly oriented along 
the $z$ direction, simple continuum 
arguments predict
that the flow is expansive and the pressure parabolic
along the channel \cite{volfsonpnas}: Assuming that the system reaches a 
steady state with constant rod density (while new rods
are created, others exit the channel), the continuity equation 
reads ${\mathbf \nabla}\cdot {\mathbf v}=a$ and can be integrated as
$v^{(0)}_z(z)=az$ and $v^{(0)}_x=0$. In the overdamped limit, the 
momentum conservation equation reads 
${\mathbf \nabla}\cdot\sigma-\mu {\mathbf v}=0$, where $\sigma$ is
the stress tensor. 
Imposing the boundary condition $\sigma=0$ at $z=\pm L$, 
one deduces that: 
\begin{equation}\label{eqpressure}
\sigma^{(0)}_{zz}(z)=\frac{1}{2}\mu a(z^2-L^2).
\end{equation}
Hence, in response to the necessary growth of the rods the pressure adopts 
a parabolic profile and is maximal at the center of the channel ($z=0$), 
where $v_z$ vanishes. As illustrated by Eq. (\ref{eqpressure}),
varying the parameter $\mu$ allows to vary the magnitude of the average
compressive load in the system.

\section{Behavior of the nematic order parameter}\label{results}

We simulated growing colonies starting from a few randomly
oriented rods distributed in the channel. 
At large times, the density is roughly constant over time 
and the channel is filled with approximately 1000 rods 
in the examples of Figure \ref{mu045}. 
To measure the degree of alignment of the rods 
we calculated the scalar nematic order parameter:
\begin{equation}
 Q= \overline{[\langle \text{cos} \ 2 \phi \rangle^2 +\langle \text{sin} 
 \ 2 \phi \rangle^2 ]^{1/2}},
 \label{parametro_orden_bidimensional}
\end{equation}
where $\phi$ is the angle between the rod axis and some reference axis (the 
channel axis $\hat{z}$).
The brackets above denote averages over all rods (spatial averaging) 
and the overbar, temporal averaging. In other words, $Q/2$ is the time 
average of the largest eigenvalue of the tensor order parameter in two 
dimensions, $\langle u_{\alpha}u_{\beta}
-\frac{1}{2}\delta_{\alpha\beta}\rangle$, where the
$u_{\alpha}$'s are the components of the orientational unit vector
of a rod \cite{doiedwards}. When the colony is in the disordered state $Q$ 
is close to zero, while perfect nematic order corresponds to $Q=1$.

\subsection{Effects of rod shape and of friction}

\begin{figure}
\begin{center}
\includegraphics[width=0.51 \textwidth]{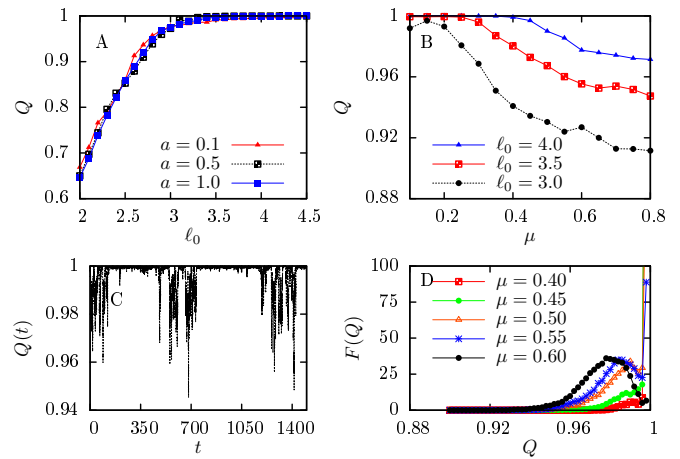}
\caption{(Color online) {\bf A}: Mean order parameter $Q$ 
as a function of the mean length of the rods at division 
($\mu=0$, $\nu_{cc}=0$). 
The growth rate $a$ has a little impact on $Q$, as shown by
the 3 overlapping curves.  
{\bf B}:  Mean order parameter $Q$ as a function of $\mu$ for
different mean rod length ($\nu_{cc}=0$). 
{\bf C}: A time series $Q(t)$ 
showing intermittent behavior at $\mu=0.45$, $\ell_0=4$. 
{\bf D}: Probability distribution 
function of the global order parameter $Q(t)$ for different friction 
coefficients $\mu$. As $\mu$ increases, the most probable $Q(t)$ takes
lower values.}
\label{first_panel}
\end{center}
\end{figure}

We first consider systems with vanishing frictions ($\mu=0$, $\nu_{cc}=0$) 
and vary $\ell_0$ (Figures \ref{mu045}A-C). Even in this case where stresses
are very small, the system may not be able to order perfectly. 
Figure \ref{first_panel}A shows the nematic order parameter as a function 
of $\ell_0$. It is observed that for $\ell_0>3$ perfect nematic order is 
reached, whereas it decays rapidly if $\ell_0<3$. In systems of short rods
({\it e.g.}, $\ell_0=2$, Fig. \ref{mu045}A) many disoriented 
regions are present and persist over time (see animation in \cite{SI}).  
It is worth noting that many bacteria such as {\it E. coli} have
aspect ratio larger than $3$ \cite{turb2} and may therefore be prone 
to form dense ordered colonies in the presence of boundaries. The growth 
rate, on the other hand, has little impact on $Q$ in the asymptotic regime, 
as the three curves with $a=0.1, 0.5$ and $1.0$ collapse 
onto each other in figure \ref{first_panel}A.

When the friction $\mu$ is finite, nematic order can be significantly
lower than in the case $\mu=0$, even for colonies of long rods ($\ell_0\ge 3$).
As shown by Figure \ref{mu045}, the 
ordered states are roughly composed of flowing columns of rods
parallel to each other. A larger friction should increase the pressure 
$-\sigma_{zz}$ exerted along the channel axis and thus increase
the repulsive interaction forces between neighboring rods of a same column. 
According to the continuum 
analysis presented in \cite{physbiol} this compressive energy can 
be released if the columns of rods bend (or buckle), producing 
less ordered configurations ($Q<1$). As expected from this scenario, we
observe that $Q$ decays with $\mu$ (Figure \ref{first_panel}B). For a 
fixed $\mu$, systems with larger $\ell_0$ are more ordered. This is also in 
qualitative agreement with the prediction of \cite{physbiol}, where the bending 
constant $\xi$ in the elastic free energy of the system was estimated from 
the overlap of a rod with the rods of the neighboring columns, leading to 
$\xi\propto \ell_0^3$.

In all the following, we fix $\ell_0=4$.
Figure \ref{first_panel}C shows a typical time evolution of the order
parameter $Q(t)$, obtained by taking the space average only, at relatively
high friction ($\mu=0.45$). The colony 
can exhibit long periods of high nematic order, interrupted once in a while 
by bursts of disorder or \lq\lq turbulence" (see animation in \cite{SI}). 
This intermittent behavior of $Q(t)$ 
is observed in a relatively narrow range of frictions, $\mu=0.40-0.50$. 

\subsection{Intermittent dynamics}

Following a method similar to that proposed in ref. \cite{huepealdana} 
to characterize the intermittent dynamics of an ordered active system, 
we extract from the corresponding time series the probability distribution 
function of $Q(t)$, for different values of the friction drag. 
As shown by Figure \ref{first_panel}D, with $\mu =0.4$ (or lower) the 
distribution of $Q(t)$ is very peaked near unity, whereas with $\mu=0.6$ 
(or larger), completely ordered configurations are never reached during 
a typical simulation time. 
In the latter high friction range, the distribution has a most 
probable value $<1$ and a 
larger variance. There is an intermediate regime, roughly in the range 
$\mu\in[0.4,0.6]$, where the 
distribution is peaked at $Q=1$ and also has a second local maximum
at some value $Q_{max}<1$. We thus consider in this regime that,
at any given time, the system can be either in an \lq\lq ordered" or in
a \lq\lq disordered" phase, depending whether 
$Q(t)>Q^{\ast}$ or $Q(t)<Q^{\ast}$, respectively, where $Q^{\ast}$ is a 
crossover value. Here we choose $Q^{\ast}$ as given by the secondary maximum
$Q_{max}$ of the distribution. The results are not very 
sensitive to other choices of $Q^{\ast}$.

\begin{figure}
\begin{center}
\includegraphics[width=0.5 \textwidth]{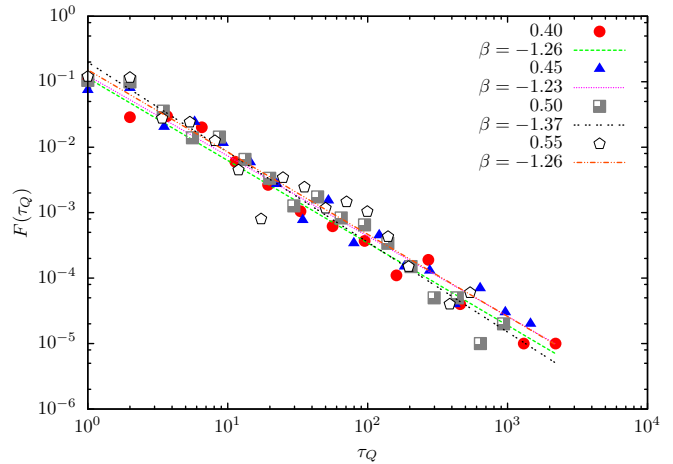}
\caption{(Color online) Probability distribution function (p.d.f.) 
for the duration $\tau_Q$ of the laminar periods, with $\mu=0.45-0.55$. 
All curves show an inverse power-law behavior. The lines are best fits
to the data, with their respective exponent estimates.}
\label{zeroth_panel}
\end{center}
\end{figure}

In this intermediate friction range, we can thus define a ordered 
(or \lq\lq laminar") time interval \cite{pomeau} as the duration $\tau_Q$ 
separating two consecutive disordered episodes. 
To measure these durations, we record the time periods during which $Q(t)$ 
remains larger than $Q^{\ast}$ without interruption. The probability 
distribution function (p.d.f.) of $\tau_Q$ is shown in Figure \ref{zeroth_panel}
and exhibits a clear inverse power-law behavior over 3 decades,
$F(\tau_Q)\sim \tau_Q^{-\beta}$. The typical value of $\beta$ is 1.2 
and depends little on $\mu$.

\section{Pressure fluctuations}

We next monitor the virial stress tensor caused
by pairwise interactions between rods and defined as
\begin{equation}
\sigma_{\alpha \beta}({\mathbf r},t)=\frac{1}{2\mathcal{V}} 
       \sum_{c, i} r_{\alpha}^{ic} F_{\beta}^{ic},
\label{eq_sigma0}       
\end{equation}
where ${\bf r}^{ic}$ is a vector from the center of mass of the rod  $i$ to a
point of contact with another rod, the index $i$ runs over all rods in
a small mesoscopic volume $\mathcal{V}$ around the position ${\mathbf r}$; 
the index $c$ runs over all points of contacts. 

\subsection{Global fluctuations}

\begin{figure}
\begin{center}
\includegraphics[width=0.52 \textwidth]{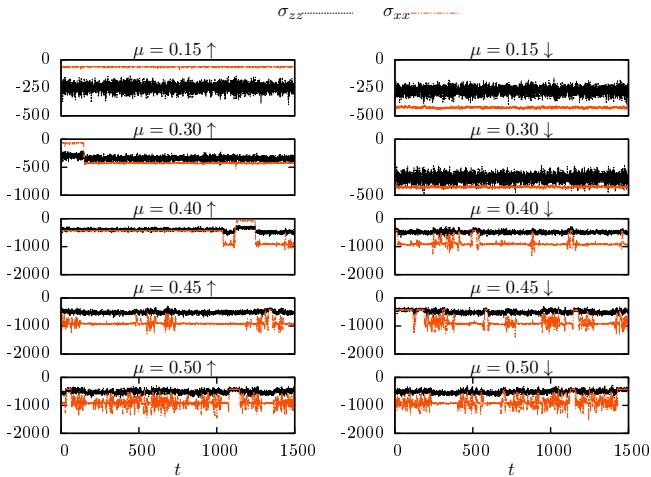}
\caption{(Color online) Time evolution of the spatially averaged 
stress tensor components 
$\langle\sigma_{xx}\rangle(t)$ (continuous orange - gray - line) and 
$\langle\sigma_{zz}\rangle(t)$ (dotted black line) for 
different values of $\mu$ during a cycle. {\it Left panel:} 
System with increasing friction ($\uparrow$ branch). 
{\it Right panel:} System with decreasing friction ($\downarrow$ branch).}
\label{spanel}
\end{center}
\end{figure}

To study the temporal fluctuations 
of the stresses in the system as a whole, we consider the spatially 
averaged stress:
\begin{equation}
\langle\sigma_{\alpha \beta}\rangle(t)=
\langle\sigma_{\alpha \beta}({\mathbf r},t)\rangle.
\end{equation}
Typical time series of $\langle\sigma_{xx}\rangle=-\langle P_{x}\rangle$ 
and $\langle\sigma_{zz}\rangle=-\langle P_{z}\rangle$ for different
values of $\mu$ are shown in Figure \ref{spanel}. In order to investigate 
possible hysteresis effects, we varied $\mu$ cyclically in a same simulation.
In the left panel of Fig. \ref{spanel}, a system is prepared
with $\mu=0.1$ and evolves during 1500 time units. The friction is then 
incremented of $0.05$ and kept constant for another 1500 time units. 
The procedure is repeated up to $\mu=0.80$ (\lq\lq $\uparrow$" branch). 
From there, the friction is decreased in a similar way with decrements 
of $0.05$ down to $0.1$ again (\lq\lq $\downarrow$" branch, right panel 
of Figure \ref{spanel}).

\begin{figure}
\begin{center}
\includegraphics[width=0.32 \textwidth]{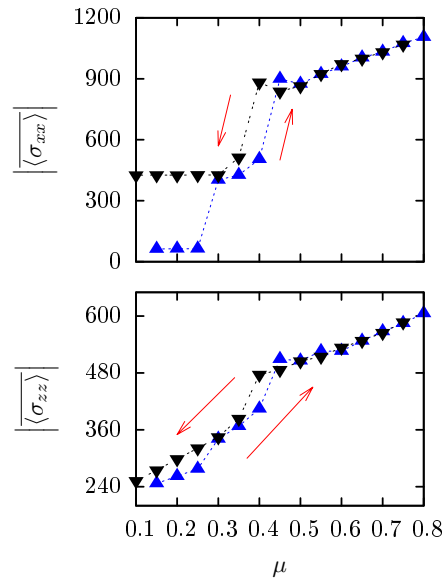}
\caption{(Color online) Time averages $\overline{\langle \sigma_{xx} \rangle}$ 
and  $\overline{\langle \sigma_{zz} \rangle}$ of the spatially averaged 
stress tensor components. The $\uparrow$ and $\downarrow$ branches of the 
cycle are labeled with blue up-triangles and 
black down-triangles, respectively.}
\label{graphS}
\end{center}
\end{figure}

The spatially averaged pressure in the longitudinal direction, 
$\langle \sigma_{zz}\rangle(t)$, fluctuates little in time and does not
show clear signs of hysteresis (see also the lower panel of Fig. \ref{graphS}). 
However, the pressure in the direction transverse to the channel axis, 
$\langle \sigma_{xx}\rangle(t)$, exhibits much larger temporal variations
(orange - gray - curves of Fig.\ref{spanel}). 
In Fig. \ref{spanel}, for $\mu=0.30$ and $0.40$ 
in the $\uparrow$ branch, for instance, one observes step-like variations 
or abrupt jumps occurring at random times between different 
stationary values. The average pressure in the $\hat{x}$ direction 
can vary in time by a factor of $\sim 7$ in a same dynamics. 

\begin{figure}
\begin{center}
\includegraphics[width=0.48 \textwidth]{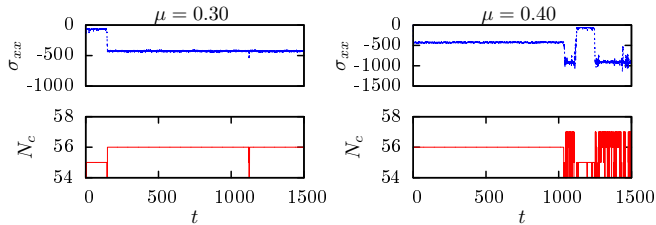}
\caption{(Color online) The spatially averaged 
stress tensor component $\langle\sigma_{xx}\rangle(t)$ (upper panels) 
at a given time is
closely related to the number $N_c$ of rod columns that fill the channel at
the same time in the transverse direction (lower panels).}
\label{columns}
\end{center}
\end{figure}

As shown by Figure \ref{columns}, these practically discrete jumps in the 
transversal pressure are due to the rapid formation (or elimination) 
of one or more columns of rods, which are oriented along the $\hat{z}$ direction.
The number of rod columns exhibits a similar step-like dynamics. 
As a new rod column appears, the system becomes more crowded in the $\hat{x}$ 
direction, resulting in a sharp increase in $|\langle \sigma_{xx}\rangle|$. 
On the contrary, when a column disappears, the pressure is relaxed.

Higher frictions cause an increase in both $P_{x}$ and $P_{z}$ on average 
(see Fig. \ref{graphS}). The increase of $P_{z}$ with $\mu$
is due to the higher friction forces exerted on the particles, has 
qualitatively predicted by
Eq. (\ref{eqpressure}). The increase in $P_{x}$ is due to the fact that
at higher frictions the system
tends to form  more columns and thus denser populations 
along the $\hat{x}$ direction. This densification was already noticed right
after the buckling instability in ref. \cite{physbiol}. 
If the friction is further decreased,
the high transverse densities may persist. For this reason, at the end 
of the hysteresis loop
($\mu=0.15\downarrow$, Fig. \ref{spanel}) the transverse pressure can be much
higher than what it was at the beginning ($\mu=0.15\uparrow$).
The higher panel of Figure \ref{graphS} illustrates the 
hysteretic behavior of the time averaged pressure 
$|\overline{\langle \sigma_{xx} \rangle}|$.
Note that at the beginning of the loop the time intervals between jumps 
can be large and thus the time averages may vary
from one simulation to another due to the limited observation time.

\subsection{Distribution of local stresses}

\begin{figure}
\begin{center}
\includegraphics[width=0.32 \textwidth]{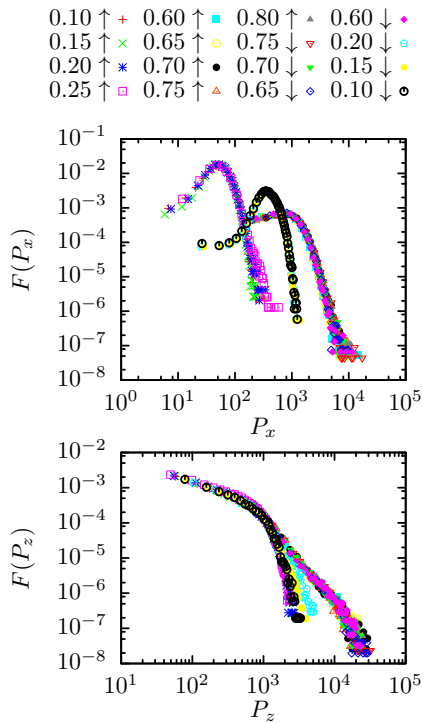}
\caption{(Color online) Probability distribution function of the local pressure 
for different values of $\mu$, which is varied cyclically:
$\mu=0.1 \rightarrow 0.8 \rightarrow 0.1$. For clarity, only the cases 
$\mu\le0.25$ and $\mu\ge0.60$ are shown. {\it Upper panel}: Distributions 
of $P_{x}$,  
where the curves for $\mu \le 0.25$ ($\uparrow$), $\mu \ge 0.60$ ($\uparrow$
and $\downarrow$) and $\mu\le 0.25$ ($\downarrow$) are located on the
left part, right part and middle part of the graph, respectively.
{\it Lower panel}: Distributions of $P_{z}$, where the curves with broader 
tails correspond to $\mu\ge0.60$.}
\label{pdfSigmapanel}
\end{center}
\end{figure}

To examine how the local pressure fluctuates in space and time, 
we display in Figure \ref{pdfSigmapanel} the probability 
distribution functions of the local stresses 
$P_{x}$ and $P_{z}$, given by Eq. (\ref{eq_sigma0}). 
These p.d.f's are obtained by aggregating all positions
and times of a given simulation.

The hysteresis effects observed above on $\langle\sigma_{xx}\rangle$, 
are noticeable in the full distribution, which has a 
characteristic scale given by its most probable value. 
In the $\uparrow$ branch,
the most probable values of $-\sigma_{xx}$
for $\mu\le0.25$ are much lower than the most probable values for $\mu\ge0.60$. 
When low friction values ($\mu\le0.25$) are reached again in the 
$\downarrow$ branch, the most probable $-\sigma_{xx}$ returns to a value larger 
than its initial value (middle curves of Figure \ref{pdfSigmapanel}, upper
panel). 

The distribution of $P_{z}$, shown in Figure \ref{pdfSigmapanel}, lower panel, 
does not exhibit such hysteresis and has a markedly 
different shape: it is monotonic decreasing and independent of $\mu$ at 
small $P_{z}$. In this regime, the distribution is approximately scale-free,
{\it i.e.}, well described by a power-law with exponent
$\approx -0.7$. Hence, the gradual increase of the average longitudinal 
stress produced by increasing $\mu$ (see Fig. \ref{graphS}) does not modify 
much how stresses are distributed locally among the rods: it only produces 
a broadening of the tail of the distribution.  Many regions carry small 
stresses and 
contribute little to the average pressure, even at high $\mu$, whereas a
few rare places have stresses much higher than average. Therefore, 
the longitudinal pressure field is very heterogeneous in space and time. 
For $\mu\le 0.25$, a fit shows that the distribution decays exponentially
at very large $P_z$. However, for $\mu\ge 0.60$, the tail of the distribution is better 
described by a second power-law, with steeper exponent $\approx -2$. 

For comparison, it is instructive to calculate the p.d.f
$F^{(0)}(P_{z})$ predicted by the continuum
theory where the rods are assumed to be perfectly aligned and where
$P_{z}(\mathbf{r},t)$ is stationary and only depends on $z$.
From the parabolic profile given by Eq.(\ref{eqpressure}) and from the 
general property $|F^{(0)}(P)dP|=|g(z)dz|$, where $g(z)$ is the distribution 
of $z$ ($g(z)=cst$ along the channel), one obtains:
\begin{equation}
F^{(0)}(P_{z})\propto \frac{1}{(1-P_{z}/P_0)^{1/2}}.
\end{equation}
According to this result, elementary regions of space where the pressure
is larger (close to the maximum $P_0$, at the center of the channel) should 
be {\it more frequent} than regions with
lower pressures. Such behavior is
opposite to that of the distributions of Figure \ref{pdfSigmapanel} 
(lower panel). This result illustrates that local disorder profoundly 
reorganizes the system by
a redistribution of stresses. This situation is reminiscent 
of the heterogeneous distributions of contact forces in static granular 
systems. Nevertheless, contact force distributions are exponential 
in static systems and thus have a typical scale \cite{forcefluct}.

\subsection{Correlations with $Q$}

\begin{figure}
\begin{center}
\includegraphics[width=0.5 \textwidth]{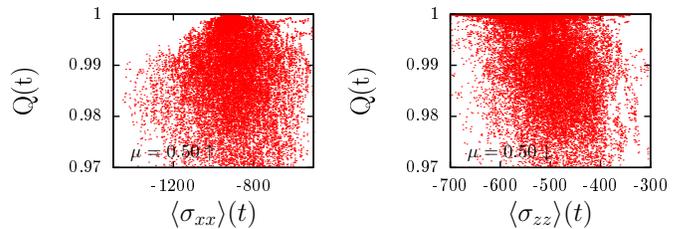}
\caption{(Color online) Parametric plots $Q(t)$ vs. 
$\langle \sigma_{xx} \rangle(t)$ and
$Q(t)$ vs. $\langle \sigma_{zz} \rangle(t)$ at $\mu=0.50$.}
\label{third_panel}
\end{center}
\end{figure}

To investigate the interplay
between nematic order and pressure at a given time, we calculated the
Pearson correlation coefficients between $Q(t)$
and  $\langle \sigma_{xx(zz)}\rangle(t)$. 
Given two arbitrary discrete time
series $a_i$ and $b_i$ of means $\overline{a}$ and $\overline{b}$, 
respectively, this coefficient is defined as
\begin{equation}
\rho_{ab}=\frac{\sum_{i=1}^{n} (a_i-\overline{a})(b_i-\overline{b})}
{\sqrt{\sum_{i=1}^n(a_i-\overline{a})^2}
\sqrt{\sum_{i=1}^n(b_i-\overline{b})^2}}.
\end{equation} 
The cases $\rho_{ab}=1$, $-1$
and $0$ correspond to perfectly correlated, anti-correlated
and not correlated variables, respectively. In Figure \ref{third_panel},
each dot represents a particular time step of a dynamics.

Table \ref{tabla} shows that
$Q(t)$ tends to be significantly anti-correlated to 
$\langle \sigma_{zz}\rangle(t)$, independently of $\mu$.
This property indicates that, at a given time, a more ordered configuration 
is likely to be subjected to larger longitudinal stresses.
This finding is consistent with the fact that gradients in the rod
orientations (mis-alignments) actually release the compressive 
longitudinal energy \cite{physbiol}. 
 
To check whether there exists a relationship between the step-like 
variations of $\langle \sigma_{xx}\rangle(t)$ at intermediate frictions
(see Figure \ref{spanel}) and the intermittent dynamics
of $Q(t)$ observed in about the same friction range, we calculated 
the Pearson coefficient between {\it (i)} $Q(t)$ and 
$\langle \sigma_{xx}\rangle(t)$,
{\it (ii)} $Q(t)$ and $|d\langle \sigma_{xx}\rangle(t)/dt|$ and {\it (iii)}
$|dQ(t)/dt|$ and $|d\langle \sigma_{xx}(t)\rangle/dt|$. As shown by
Table \ref{tabla},  very weak correlations are found in almost all cases.
Therefore, there seems to be no systematic correlations between the 
fast variations in  $\langle \sigma_{xx}\rangle(t)$ and the intermittent
bursts of nematic disorder, except maybe for lower frictions, see 
$\mu=0.4$ and $0.45$ in Table \ref{tabla}.
This suggests that the mechanisms by with the
system modulates its transversal pressure under confinement 
(through the formation or elimination of columns of growing rods)  
is not directly related to the ordering dynamics itself. The
mis-alignment of some rods does not preferentially leads to a lower
transverse pressure, contrary to what happens in the longitudinal direction.

\begin{table}
\begin{center}
\begin{tabular}{|c||c|c|c|c|}
\hline\hline
$\mu $   & $Q\ \&\ \sigma_{zz}$ & $Q\ \&\ \sigma_{xx}$  & 
          $Q\ \&\ |\dot{\sigma}_{xx}|$ & $\dot{Q}\ \&\ |\dot{\sigma}_{xx}|$ \\
\hline
    0.40  &  -0.27  & -0.17  & -0.03 & $\sim0$ \\ \hline
    0.45  &  -0.38  & -0.20  &  0.03 & $\sim0$ \\ \hline
    0.50  &  -0.34  &  0.04  & -0.01 & $\sim0$ \\ \hline
    0.55  &  -0.34  &  0.04  & -0.00 & $\sim0$ \\ \hline\hline
\end{tabular}
\caption{Numerical values for the Pearson correlation coefficient
from time series where $Q(t)$ has an intermittent behavior. These values 
correspond to the $\downarrow$ part of the cycle (similar values
are found for the $\uparrow$ part).}
\label{tabla}
\end{center}
\end{table}


\section{Conclusions}\label{conclusions}

We have studied with molecular dynamics simulations the ordering
of systems of growing elongated particles confined in a channel. 
We find that
the average nematic order parameter depends 
crucially on the rod aspect ratio, a parameter which is difficult to 
incorporate in continuum theories. Colonies fail to order parallel to
the side walls if $\ell_0<3$, even when the drag friction is vanishing. 
For $\ell_0>3$ and
at finite friction, intermittent bursts of disorder can take place and the 
periods during which the system remains well-ordered are power-law 
distributed. In another context, intermittent dynamics for the global
order parameter have already been observed in active systems of self-propelled
particles governed by the Vicsek model rules \cite{huepealdana}. 

Our results also show that the stress tensor is very anisotropic and that
the pressure field has markedly different properties 
in the directions transverse and longitudinal to the channel axis 
($\hat{x}$ and $\hat{z}$, respectively).
Whereas $P_{x}$ is relatively homogeneously distributed in space,
its spatial average can vary very rapidly in time due 
to stochastic and abrupt density variations in the lateral direction. 
This density can remain constant for long periods of time at low friction, 
which leads to hysteresis effects \cite{ohta}. 
The fast variations of the spatial average $\langle P_x\rangle(t)$ 
at intermediate frictions
do not seem to be correlated to the intermittent dynamics of the nematic 
order parameter. Comparatively, the spatially averaged 
$\langle P_z\rangle(t)$ has a much smoother behavior in time and does not 
present hysteresis,
but it is correlated to the global nematic order parameter. This is to be
expected from theoretical arguments that predict that longitudinal stresses 
should be released in systems of misaligned rods \cite{physbiol}. 

We emphasize that,
unlike $P_x$, the longitudinal pressure $P_z$ is very heterogeneously 
distributed in space, in such a way that most of the rods are subjected 
to small stresses while
very large stresses are supported by a few rods. This trend
is opposite to the prediction of a simple continuum theory (that ignores
granularity), which is that not-so-stressed rods should be less numerous than
highly stressed rods.
The distribution of $P_z$ is well fitted by a truncated power-law
at low friction and by two power-laws at large friction. 
For comparison, the probability distribution function of the 
contact forces in jammed packings of non-active grains is generically 
exponential, {\it i.e.} comparatively much more homogeneous \cite{forcefluct}.
Contact forces also remain exponentially distributed in sheared packings
of elongated particles \cite{azema}.

In the proliferating systems studied here, a global state of compressive 
stress thus emerges from individual cell growth and division. This parallels 
the case of advancing sheets of epithelial cells in a channel, where global 
states of tensile stress have been observed in experiments \cite{nphys2010}. 
In those experiments, traction did not result from leader cells at the edge 
of the sheet dragging those behind, but from the cells located deep inside 
the tissue. It was observed that the traction force also followed a profile 
approximately parabolic, and exhibited, at a fixed location, large temporal 
fluctuations. These fluctuations were exponentially distributed, though, as
in static granular materials \cite{nphys2010}.

Previous studies have shown that dense colonies of growing 
bacteria are able to self-organize and form crowds that efficiently
escape from confining domains \cite{groisman}.
Our results further suggests that active systems subjected to 
external perturbations (such as an average pressure increase) could
have the ability to self-organize in such a way that only 
a few particles would actually be affected by the perturbation. Our findings 
could have implications for understanding the growth of real
tissues and biofilms, where individual cells subjected to large 
stresses are known to grow at a reduced rate or not to grow 
at all \cite{tumorjtb,volfsonpnas}.
Colonies of bacteria or other cell types may be able to keep 
growing in adverse conditions and the study of such robustness should 
motivate further studies.

\acknowledgments

SOF acknowledges financial support from CONACYT scholarship grant 174695.
We thank V. Romero, E. Ru\'\i z-Guti\'errez, L.S. Tsimring, W. Mather and 
R. Zenit for valuable discussions.


\begin{thebibliography}{99}
%
\bibitem{libchaber} X.-L. Wu and A. Libchaber, 
Phys. Rev. Lett. {\bf 84}, 3017 (2000).
%
\bibitem{couzin} T. S. Deisboeck and I. D. Couzin,
BioEssays {\bf 31}, 190 (2009).
%
\bibitem{volfsonprl} A. Kudrolli, G. Lumay, D. Volfson, and 
L. S. Tsimring, Phys. Rev. Lett. {\bf 100}, 058001 (2008).
%
\bibitem{joanny} M.C. Marchetti, J.F. Joanny, S. Ramaswamy, T.B. 
Liverpool, J. Prost, Madan Rao, and R. Aditi Simha,
arXiv:1207.2929 [cond-mat.soft] (2012).
%
\bibitem{turb1} H. H. Wensink and H. L$\ddot{\rm o}$wen,
J. Phys.: Condens. Matter {\bf 24}, 464130 (2012).
%
\bibitem{turb2} H. H. Wensink, J. Dunkel, S. Heidenreich, K. Drescher,
R. E. Goldstein, and J. M. Yeomans, Proc. Natl. Acad. Sci. USA {109},
14308 (2012).
%
\bibitem{turb3} J. Dunkel, S. Heidenreich, K. Drescher, H. H. Wensink, 
M. Bar, and R. E. Goldstein,  Phys. Rev. Lett. {\bf 110}, 228102 (2013).
%
\bibitem{jamming} S. Henkes, Y. Fily, and  M. C. Marchetti,
Phys. Rev. E {\bf 84}, 040301(R) (2011).
%
\bibitem{swinney} H. P. Zhang, A. Be'er, E.-L. Florin, and H. L. Swinney,
Proc. Natl. Acad. Sci. USA {\bf 107}, 13626 (2010).
%
\bibitem{peruaniprl} F. Peruani, J. Starru\ss, V. Jakovljevic, L. 
S\o gaard-Andersen, A. Deutsch, and Markus Bar,
Phys. Rev. Lett. {\bf 108}, 098102 (2012).
%
\bibitem{gianttheor} Y. Yang, V. Marceau, and G. Gompper,
Phys. Rev. E {\bf 82}, 031904 2010.
%
\bibitem{hydra} R. D. Campbell, J. Morphol. {\bf 121}, 19 (1967).
%
\bibitem{tissuebiophys} G. Cheng, B. B. Youssef, P. Markenscoff, 
and K. Zygourakis, Biophys. J. {\bf 90}, 713 (2006).
%
\bibitem{tissuerepair} M. Poujade, E. Grasland-Mongrain, A. Hertzog, 
J. Jouanneau, P. Chavrier, B. Ladoux, A. Buguin, and P. Silberzan,
Proc. Natl. Acad. Sci. USA {\bf 104}, 15988 (2007).
%
\bibitem{tumorjtb} A. R. Kansal, S. Torquato, G. R. Harsh, E. A.
Chiocca, and T. S. Deisboeck, J. Theor. Biol. {\bf 203}, 367 (2000).
%
\bibitem{baskaranprl2008} A. Baskaran and M. C. Marchetti,
Phys. Rev. Lett. {\bf 101}, 268101 (2008).
%
\bibitem{baskaranpre2008} A. Baskaran and M. C. Marchetti,
Phys. Rev. E {\bf 77}, 011920 (2008).
%
\bibitem{chate} F. Ginelli, F. Peruani, M. B$\ddot{\rm a}$r, and H. Chat\'e,
Phys. Rev. Lett. {\bf 104}, 184502 (2010).
%
\bibitem{aranson} A. Peshkov, I. S. Aranson, E. Bertin, H. Chat\'e, 
and F. Ginelli,  Phys. Rev. Lett. {\bf 109}, 268701 (2012).  
%
\bibitem{bartolo} T. Brotto, J.-B. Caussin, E. Lauga, and D. Bartolo,
Phys. Rev. Lett. {\bf 110}, 038101 (2013).
%
\bibitem{volfsonpnas} D. Volfson, S. Cookson, J. Hasty, and L. S. Tsimring,
Proc. Natl. Acad. Sci. USA {\bf 105}, 15346 (2008).
%
\bibitem{physbiol} D. Boyer, W. Mather, O. Mondrag\'on-Palomino, 
S. Orozco-Fuentes, T. Danino, J. Hasty, and L. S. Tsimring,
Phys. Biol. {\bf 8}, 026008 (2011).
%
\bibitem{volfsonkudrolli}   D. Volfson, A. Kudrolli, and L. S. Tsimring,
Phys. Rev. E {\bf 70}, 051312 (2004).
%
\bibitem{groisman} H. Cho, H. J$\ddot{\rm o}$nsson, K. Campbell, P. Melke, 
J. W. Williams, B. Jedynak, A. M. Stevens, A. Groisman, A. Levchenko,
PLoS Biol. {\bf 5}, e302 (2007).
%
\bibitem{doiedwards} M. Doi and S. F. Edwards,{\it The theory of
polymer dynamics} (Oxford University Press, Oxford, 1986).
%
\bibitem{SI} Supplemental Material files.
%
\bibitem{huepealdana} C. Huepe and M. Aldana,
Phys. Rev. Lett. {\bf 92}, 168701 (2004).
%
\bibitem{pomeau} P. Berge, Y. Pomeau, and C. Vidal, 
{\it Order Within Chaos: Towards a Deterministic Approach to Turbulence} 
(John Wiley and Sons, Inc., New York, 1987).
%
\bibitem{forcefluct} C.-h. Liu, S. R. Nagel, D. A. Schecter, S. N. Coppersmith,
S. Majumdar, O. Narayan, and T. A. Witten, Science {\bf 269}, 513 (1995).
%
\bibitem{azema} E. Az\'ema and F. Radja$\ddot{\rm \i}$,
Phys. Rev. E {\bf 85}, 031303 (2012).
%
\bibitem{ohta} Hysteresis behavior is also found 
in (non-confined) systems of deformable self-propelled particles
with repulsive interactions, see Y. Itino, T. Ohkuma, and T. Ohta,
J. Phys. Soc. Jpn. {\bf 80}, 033001 (2011). 
%
\bibitem{nphys2010} X. Trepat, M. R. Wasserman, T. E. Angelini, E. Millet, 
D. A. Weitz, J. P. Butler, and J. J. Fredberg,
Nature Phys. {\bf 5}, 426 (2010).
%
\end{thebibliography}
\end{document}